\documentclass[11pt]{article}

\usepackage[margin=1.1in]{geometry}
\usepackage{graphicx}

\newcommand{\bff}{{\bf f}}
\newcommand{\bfb}{{\bf b}}
\newcommand{\bfA}{{\bf A}}
\newcommand{\x}{{\bf x}}
\newcommand{\z}{{\bf z}}
\newcommand{\zero}{{\bf 0}}

\begin{document}


\title{Parallel Software to Offset 
the Cost of Higher Precision\thanks{Supported by 
the National Science Foundation under grant DMS 1854513.}}

\author{Jan Verschelde\thanks{University of Illinois at Chicago,
Department of Mathematics, Statistics, and Computer Science,
851 S. Morgan St. (m/c 249), Chicago, IL 60607-7045
Email: {\tt janv@uic.edu}, URL: {\tt http://www.math.uic.edu/$\sim$jan}.}}

\date{11 December 2020}


\maketitle

\begin{abstract}
Hardware double precision is often insufficient to solve
large scientific problems accurately.
Computing in higher precision defined by software causes
significant computational overhead.
The application of parallel algorithms compensates for this overhead.
Newton's method to develop power series expansions of algebraic space curves
is the use case for this application.
\end{abstract}

\subsection*{1. Problem Statement and Overview}

While parallel computers are fast and can solve large problems,
the propagation of roundoff errors increases as problems grow larger
and the hardware supports only double precision.  
If we can afford the same time as on a sequential run, 
then we ask for {\em quality up}:
by how much can we improve the quality of the results with a parallel run?
To us, quality means accuracy.  The goal is to compensate for the
overhead of multiple double arithmetic with parallel computations.

The focus of this paper is on recently developed code for new algorithms
described in~\cite{BV18}, \cite{TVV20a,TVV20b}, added to PHCpack~\cite{Ver99}.
PHCpack is a free and open source package to apply Polynomial Homotopy 
Continuation to solve systems of many polynomials in several variables.
Continuation methods are classic algorithms in applied mathematics,
see e.g.~\cite{Mor87}.
Ada is the main language in which the algorithms in PHCpack have been
developed during the past thirty years.  
Strong typing and standardization make that the same code runs on different
platforms (Linux, Windows, Mac OS X) and that the same code continues to run,
even after decades, without the need to update for upgrades of the language.
Ada tasking provides an effective high level tool to develop algorithms
for parallel shared memory computers; 
see~\cite{BW07} or~\cite{MSH11} for introductions.

Using QDlib~\cite{HLB01} and the software CAMPARY~\cite{JMPT16}, we extend
the range of precision offered by hardware doubles~\cite{MBDJJLMRT18},
as a step towards rigorous verification.
In our numerical study of algebraic curves~\cite{Wal50}, we apply algorithmic
differentiation~\cite{GW08}, numerical linear algebra~\cite{GV83},
and rational approximation techniques~\cite{BG96}.

The first three sections in this paper motivate the need for higher precision
and describe the computational cost overhead.
This overhead then motivates the application of multitasking.
All computational experiments for this paper were done on
a CentOS Linux workstation with 256 GB RAM
and two 22-core 2.2 GHz Intel Xeon E5-2699 processors.

\newpage 

\subsection*{2. Multiple Double Numbers}

A double double is an unevaluated sum of two hardware doubles.
With the application of basic arithmetical operations in IEEE double format,
we obtain more accurate results, up to twice the accuracy of the 
hardware double precision.
In~\cite{Rum10}, double double arithmetic is described in the context
of error-free transformations; see also~\cite[Chapter~14]{MBDJJLMRT18}.
Double double and quad double arithmetic is provided by QDlib~\cite{HLB01}.
Code generators for general multiple double and multiple float arithmetical
operations are available in the software CAMPARY~\cite{JMPT16}.

As an illustration of multiple double arithmetic, consider the
computation of the 2-norm of a vectors of dimension~64 of
complex numbers generated as $\cos(\theta) + \sin(\theta) \sqrt{-1}$,
for random angles~$\theta$.
The 2-norm equals~8.
Observe the second double of the multiple double 2-norm.
\begin{verbatim}
      double double : 8.00000000000000E+00 - 6.47112461314111E-32
      triple double : 8.00000000000000E+00 + 1.78941597340672E-48
        quad double : 8.00000000000000E+00 + 3.20475411419393E-65
       penta double : 8.00000000000000E+00 + 2.24021706293649E-81
        octo double : 8.00000000000000E+00 - 9.72609915198313E-129
        deca double : 8.00000000000000E+00 + 3.05130075600701E-161
\end{verbatim}
The format of the result above is for this experiment preferable over
the decimal expansion which may appear as $7.999\ldots9$.
The Ada code for multiple double precision is available
in the free and open source software PHCpack~\cite{Ver99},
under version control at github.

Table~\ref{tabopscount} shows the cost of the basic operations
in multiple double precision, expressed in the number of
hardware double arithmetical operations.
A tenfold increase in precision from double to deca double
leads to a more than thousandfold increase in the count of
the arithmetical operations.  

\begin{table}[hbt]
\begin{center}
\begin{tabular}{c|rrrr|rrrr|rrrr}
& \multicolumn{4}{c|}{double double}
& \multicolumn{4}{c|}{triple double}
& \multicolumn{4}{c}{quad double} \\
& \multicolumn{1}{c}{$+$} & \multicolumn{1}{c}{$-$}
& \multicolumn{1}{c}{$*$} & \multicolumn{1}{c|}{$/$}
& \multicolumn{1}{c}{$+$} & \multicolumn{1}{c}{$-$}
& \multicolumn{1}{c}{$*$} & \multicolumn{1}{c|}{$/$}
& \multicolumn{1}{c}{$+$} & \multicolumn{1}{c}{$-$}
& \multicolumn{1}{c}{$*$} & \multicolumn{1}{c}{$/$} \\ \hline
add &   8 &  12 &  & 
    &  13 &  22 &  &
    &  35 &  54 &  & \\
mul &   5 &   9 &  9 &
    &  83 &  84 &  42 &
    &  99 & 164 &  73 & \\
div &  33 &  18 &  16 & 3
    & 113 & 214 &  63 & 4 
    & 266 & 510 & 112 & 5 \\ \hline \hline
& \multicolumn{4}{c|}{penta double}
& \multicolumn{4}{c|}{octo double}
& \multicolumn{4}{c}{deca double} \\
& \multicolumn{1}{c}{$+$} & \multicolumn{1}{c}{$-$}
& \multicolumn{1}{c}{$*$} & \multicolumn{1}{c|}{$/$}
& \multicolumn{1}{c}{$+$} & \multicolumn{1}{c}{$-$}
& \multicolumn{1}{c}{$*$} & \multicolumn{1}{c|}{$/$}
& \multicolumn{1}{c}{$+$} & \multicolumn{1}{c}{$-$}
& \multicolumn{1}{c}{$*$} & \multicolumn{1}{c}{$/$} \\ \hline
add &   44 &   78 & & 
    &   95 &  174 & & 
    &  139 &  258 & & \\
mul &  162 &  283 & 109 &
    &  529 &  954 & 259 &
    &  952 & 1743 & 394 & \\
div &  474 &  898 & 175 & 6
    & 1599 & 3070 & 448 & 9
    & 2899 & 5598 & 700 & 11
\end{tabular}
\caption{Number of double operations for addition (add),
multiplication (mul), division (div), required for a 2-fold, 
3-fold, 4-fold, 5-fold, 8-fold, and 10-fold increase in precision.}
\label{tabopscount}
\end{center}
\end{table}

The operation counts in Table~\ref{tabopscount} then motivate the
need for parallel computations as follows.
What takes a millisecond to compute in double precision will take
several seconds in deca double precision.  
A program that finishes in a second in double precision will take more
than an hour in deca double precision.  
A computation in double precision that takes a hour will 
in deca double precision take more than a month to finish.

\newpage

\subsection*{3. Polynomials as Truncated Power Series}

Writing a polynomial backwards (starting at the constant term
and then listing the monomials in the increasing degree order),
leads to the interpretation of a polynomial as the sum of the
leading terms in a power series.  Unlike polynomials,
every power series with a leading nonzero constant term has an inverse.
One can divide power series by another and calculate with power series 
similar as to number arithmetic~\cite{Wal50}.
In this section, we consider Newton's method where the arithmetic
happens with truncated power series instead of with ordinary numbers.

One common parameter representation for points on the circle with
radius one is $(\cos(t), \sin(t))$, for $t \in [0, 2\pi[$.
With truncated power series arithmetic we can approximate this
representation.  Consider a system of two polynomials in three variables:
\begin{displaymath}
 \left\{
  \begin{array}{l}
     t - 1/6 t^3 + 1/120 t^5 - 1/5040 t^7 - y = 0 \\
     x^2 + y^2 - 1 = 0.
  \end{array}
 \right.
\end{displaymath}
The first polynomial represents the equation
$ y = t - 1/6 t^3 + 1/120 t^5 - 1/5040 t^7$.
The right hand side of this equation contains the first four
leading terms of the Taylor expansion of $\sin(t)$.

Given the leading terms of $\sin(t)$, running Newton's method,
with 8 as the truncation degree of the power series,
starting at $x = 1$, $y = 0$, and $t = 0$,
the leading terms of $\cos(t)$ will appear
as the solution series for~$x$.
Indeed, the numerical output contains
\begin{verbatim}
      2.48015873015868E-05*t^8 - 1.38888888888889E-03*t^6
      + 4.16666666666667E-02*t^4 - 5.00000000000000E-01*t^2 + 1.
\end{verbatim}
The second polynomial has floating-point coefficients which
approximate the Taylor series of the $\cos(t)$, in particular
$x = 1 - 1/2 t^2 + 1/24 t^4 - 1/720 t^6 + 1/40320 t^8$.
Although many programmers will experience the temptation to
display {\tt 5.00000000000000E-01} as {\tt 1/2}, 
the {\tt 7} in the number {\tt 4.16666666666667E-02} 
gives an indication about the size of the roundoff error.
This information would be lost if one would display the result
by the nearest rational number~{\tt 1/24}.

Looking at polynomials as truncated power series has the benefit
that the solver can handle larger classes of nonlinear systems,
as the first equation of the above polynomial system can be viewed
as an approximation for $\sin(t) - y = 0$.
With truncated power series as coefficients, the solutions of systems 
where the number of variables is one more than the number of equations
are also power series.
Although the convergence radius of power series can be limited,
power series serve as input to compute highly accurate rational
approximations for functions~\cite{BG96}.

Even as the above calculation was performed in double precision,
Newton's method did not run on vectors of numbers,
but on vectors of truncated power series,
represented as power series with vector coefficients.
Working with truncated power series causes an extra cost overhead
and provides an additional motivation for parallel computations.
In particular, the multiplication of two power series truncated
to degree~$d$ requires $(d+2)(d+1)/2$ multiplications and
$(d+1)d/2$ additions.  For a modest degree $d = 8$,
the formulas in the previous sentence evaluate to 45 and~36.
For $d = 32$ the corresponding numbers are 561 and~528.
These numbers predict the cost overhead factors in working
with truncated power series arithmetic.

Working with power series of increasing degrees of truncation
leads to more roundoff and requires therefore arithmetic in
higher precision, as will be made explicit in the next section.

\newpage

\subsection*{4. Newton's Method on Truncated Power Series}

In this section we make our problem statement more precise.
In particular, running a lower triangular block Toeplitz solver
results in a loss of accuracy.

One step of Newton's method requires
evaluation and differentiation of the system,
followed by the solution of a linear system.
Consider $\bff(\x) = \zero$ as a system of polynomials in several variables,
with coefficients as truncated power series in the variable~$t$,
where $\bff = (f_1,f_2, \ldots, f_N)$, $\x = (x_1,x_2, \ldots,x_n)$,
and $N \geq n$.  For $N > n$, the linear systems are solved in the
least squares sense, either with QR or SVD;
for $N = n$, an LU factorization can be applied;
see~\cite{GV83} for an introduction to matrix factorizations.

Then we compute $\x(t)$ a power series solution to $\bff(\x) = \zero$,
starting at a point $\x(0) = \z$, $\x(t) = \z + \x_1 t + \x_2 t^2 + \cdots$.
With linearization~\cite{BV18},
instead of vectors and matrices of power series,
we consider power series with vectors and matrices as coefficients.
A matrix is denoted with a capitalized letter, e.g.: $A$;
vectors are denoted in bold, e.g.: $\x$, $\bfb$. 
To compute the update $\Delta \x$ to the solution in Newton's method,
a linear system is solved.  With truncated power series arithmetic,
this linear system is written in short as $\bfA(t) \Delta \x(t) = \bfb(t)$.
The given coefficients in this equation $\bfA$ and $\bfb$,
where $\bfA$ is a vector of matrices $\bfA = (A_0, A_1, \ldots, A_d)$
and $\bfb$ is a vector of power series.


In linearized format, for truncation degree~$d$,
$\bfA(t) \Delta \x(t) = \bfb(t)$ represents
\begin{displaymath}
  \begin{array}{ll}
   & {\displaystyle 
         \left( A_0 + A_1 t + A_2 t^2 + \cdots + A_d t^d \right)}
     {\displaystyle
         \left( \Delta \x_0 + \Delta \x_1 t + \Delta \x_2 t^2
                + \cdots + \Delta \x_d t^d \right)} \\
  = & {\displaystyle 
         \bfb_0 + \bfb_1 t + \bfb_2 t^2 + \cdots + \bfb_d t^d }.
\end{array}
\end{displaymath}
The $A_0$ is the matrix of all partial derivatives 
of the polynomials in~$\bff$ at the leading constant coefficient
of the power series expansion of the solution vector.
Methods of algorithmic differentiation~\cite{GW08} lead to
an efficient calculation of~$\bfA(t)$.
In particular, computing all $n$ partial derivatives of a function $f$ in
$n$ variables requires about 3 (and not $n$) times the cost to evaluate~$f$.

Expanding the multiplication and rearranging the terms according
to the powers of~$t$ leads to a lower triangular block system:
\begin{displaymath}
  \left[
   \begin{array}{ccccc}
      A_0 & & & & \\
      A_1 & A_0 & & & \\
      A_2 & A_1 & A_0 & & \\
      \vdots & \vdots & \vdots & \ddots & \\
      A_d & A_{d-1} & A_{d-2} & \cdots & A_0
   \end{array}
  \right]
  \left[
     \begin{array}{c}
        \Delta \x_0 \\ \Delta \x_1 \\ \Delta \x_2 \\ \vdots \\ \Delta \x_d
     \end{array}
  \right]
  =
  \left[
    \begin{array}{c}
       \bfb_0 \\ \bfb_1 \\ \bfb_2 \\ \vdots \\ \bfb_d
    \end{array}
\right].
\end{displaymath}
Forward substitution is applied as follows.
First solve $A_0 \Delta \x_0 = \bfb_0$.
Once $\Delta \x_0$ is known,
The second equation $A_1 \Delta \x_0 + A_0 \Delta \x_1 = \bfb_1$
then becomes $A_0 \Delta \x_1 = \bfb_1 - A_1 \Delta \x_0$.
After the computation of $\Delta \x_1$, the third equation
$A_2 \Delta \x_0 + A_1 \Delta \x_1 + A_0 \Delta \x_2 = \bfb_2$ turns
into the equation
$A_0 \Delta \x_2 = \bfb_2 - A_2 \Delta \x_0 - A_1 \Delta \x_1$, etc.

The exploitation of the block structure reduces the computation of $d$
linear systems with the same $N$-by-$n$ coefficient matrix~$A_0$.
Suppose that in each step up to two decimal places of accuracy would be lost,
then the accuracy loss of the last $\Delta \x_d$ could be as large as~$2d$.
Even with a modest degree of~8, in double precision, this would imply that
all 16 decimal places of accuracy are lost.
A loss of 16 decimal places of accuracy in double double precision still
leads to sufficiently accurate results.
This argument is expressed formally in~\cite{TVV20b}.

\newpage

\subsection*{5. Application of Multitasking}

In multiple double arithmetic, programs become compute bound,
which is beneficial on computers with faster processor speed
than memory speed.  On a parallel shared memory computer,
multiple threads run within one process.
In the type of multitasking applied in this paper,
each task is mapped to one kernel thread.
Typically the total number of tasks in each parallel run should 
then not exceed (twice) the number of available cores on the processor. 

Current processors run at the speed of a couple of GHz and have thus
a theoretical peak performance of one billion floating-point operations
per second.  One billion is $10^9$ or $1,000 \times 1,000 \times 1,000$.
The first two thousands in this billion represent roughly the overhead
caused by multiple double and power series arithmetic, with the last
thousand the original computational cost in double precision.
This rough estimate explains that one job will typically take
several seconds and is relatively much larger than the cost of
launching several threads.

\begin{figure}[hbt]
\begin{center}
\begin{picture}(400,30)
\put(8,10){queue}
\put(45,8){\line(1,0){80}}
\put(45,18){\line(1,0){80}}
\put(45,8){\line(0,1){10}}  \put(55,8){\line(0,1){10}}
\put(65,8){\line(0,1){10}}  \put(75,8){\line(0,1){10}}
\put(85,8){\line(0,1){10}}  \put(95,8){\line(0,1){10}}
\put(105,8){\line(0,1){10}} \put(115,8){\line(0,1){10}}
\put(125,8){\line(0,1){10}}
\put(50,13){\circle*{3}}  \put(50,13){\vector(0,-1){15}}  \put(48,23){${}_1$}
\put(60,13){\circle*{3}}  \put(60,13){\vector(0,-1){15}}  \put(58,23){${}_2$}
\put(70,13){\circle*{3}}  \put(70,13){\vector(0,-1){15}}  \put(68,23){${}_3$}
\put(80,13){\circle*{3}}  \put(80,13){\vector(0,-1){15}}  \put(78,23){${}_4$}
\put(90,13){\circle*{3}}  \put(90,13){\vector(0,-1){15}}  \put(88,23){${}_5$}
\put(100,13){\circle*{3}} \put(100,13){\vector(0,-1){15}} \put(98,23){${}_6$}
\put(110,13){\circle*{3}} \put(110,13){\vector(0,-1){15}} \put(108,23){${}_7$}
\put(120,13){\circle*{3}} \put(120,13){\vector(0,-1){15}} \put(118,23){${}_8$}
\put(155,10){counter}
\put(200,8){\line(1,0){10}}  \put(200,8){\line(0,1){10}}
\put(200,18){\line(1,0){10}} \put(210,8){\line(0,1){10}}
\put(203,12){${}_4$}
\put(220,10){guarded by a binary semaphore}
\end{picture}
\caption{Schematic of a job queue 
         with a counter guarded by a binary semaphore.}
\label{figjobqueue}
\end{center}
\end{figure}
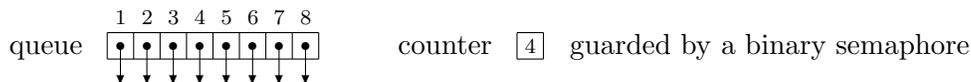

All data is allocated and defined before the threads are launched.
Figure~\ref{figjobqueue} illustrates the design of a job queue
with 8 jobs and 4 finished jobs, as the counter counts
the number of finished jobs.  The arrows in the picture point
to the read only input data and the work space for each job. 
While the input may point to shared data, each job has distinct,
non overlapping memory locations for the work space needed for each job.
When a task needs to work on the next job, it requests entry to the
binary semaphore that guards the counter for the next job. 
Once entry is granted, the tasks increments the value of the counter
and releases the lock.  The time spent inside a critical section is
thus minimal.

The organization of all computational work into a job queue
determines the granularity of the parallelism.
In the application of running Newton's method,
medium grained parallelism is applied.
For the evaluation and differentiation of a system of polynomials,
one job is concerned with one polynomial.
For the solving of the block triangular linear system,
one job is the update of one right hand side vector after
the computation of one update vector.
The solution of the linear system can happen only after all
polynomials in the system are evaluated and differentiated.
The synchronization between those two stages is performed
by terminating all tasks and launching a new set of tasks
for the next stage.  Details are described in~\cite{TVV20b}.

The main executable in PHCpack is {\tt phc}.  The user can
specify the number of tasks at the command line,
e.g., call the solver with eight tasks as {\tt phc -b -t8}.
If no number follows the {\tt -t}, then the number of tasks equals
the number of available kernel threads.
Below is the output of {\tt time phc -b} and {\tt time phc -b2 -t},
respectively in double and double double precision,
on the cyclic 7-roots benchmark, using 88 threads.
\vspace{-2mm}
\begin{verbatim}
             real    0m10.310s         real    0m1.661s   
             user    0m10.188s         user    1m12.226s 
             sys     0m0.008s          sys     0m0.083s    
\end{verbatim}
\vspace{-2mm}
The numbers after {\tt real} are the elapsed wall clock time.
With multitasking, the speedup in double double
over double precision is $10.310/1.661 \approx 6.2$.
We have speedup and quality up.

\newpage

\subsection*{6. A Numerical Experiment using Multiple Doubles}

At the end of~\cite{TVV20b}, we reported an instance
where quad double precision was insufficient for Newton's method to converge
and compute the coefficients of the series past degree~15.

Details for the experiment can be found in~\cite{TVV20b},
a short summary follows.
The series development start at a generic point on a 7-dimensional surface
of cyclic 128-roots, defined by a polynomial system of 128 polynomials
in 128 variables, augmented with seven linear equations.
To every equation in the system, a parameter $t$ is added.
At $t=0$, the generic point on the 7-dimensional surface is then
the leading coefficient vector of the power series expansion
of the solution curve in~$t$.

For this problem, the inverse of the condition number of the matrix~$A_0$
is estimated at $4.6\mathrm{E}-6$, which implies that up to six decimal
places of accuracy may be lost in the computation of the next term of
the power series.  The accuracy of the power series is measured by
$\| \Delta \x \|$, the modulus of the update to the last coefficient
in the power series.  The tolerance on $\| \Delta \x \|$ for all runs
is set to $1.0\mathrm{E}{-32}$.  Newton's method stops when 
$\| \Delta \x \| \leq 1.0\mathrm{E}{-32}$ or when the number of steps
has exceeded the maximum number of iterations.
The maximum number of iterations with Newton's method is as many as
as 8, 8, 12, and 16, for the respective degrees 8, 16, 24, and 32
of the power series. 

Table~\ref{tabcyc128} summarizes the data of the numerical experiment
with {\tt phc -u -t}.
Once $\| \Delta \x \|$ is too large for one degree,
computations for the next degree are not done.

\begin{table}[hbt]
\begin{center}
\begin{tabular}{cc||c|c|c|c}
precision & output
      & degree 8 & degree 16 & degree 24 & degree 32 \\ \hline \hline
quad  & $\| \Delta \x \|$ & $2.2\mathrm{E}{-30}$ & $1.6\mathrm{E}{+3}$ &  \\
      & \#iterations & 8 & 8 & & \\
      & seconds & 56 & 168 & & \\ \hline
penta & $\| \Delta \x \|$ & $1.1\mathrm{E}{-47}$ & $1.1\mathrm{E}{-14}$
      & $4.1\mathrm{E}{+19}$ & \\
      & \#iterations & 5 & 8 & 12 & \\
      & seconds & 46 & 231 & 722 & \\ \hline
octo  & $\| \Delta \x \|$ & $1.4\mathrm{E}{-69}$ & $9.5\mathrm{E}{-63}$
      & $3.8\mathrm{E}{-30}$ & $3.4\mathrm{E}{+3}$ \\
      & \#iterations & 5 & 6 & 12 & 16 \\
      & seconds & 128 & 472 & 1,934 & 4,400 \\ \hline
deca  & $\| \Delta \x \|$ & $1.4\mathrm{E}{-69}$ & $2.4\mathrm{E}{-95}$
      & $1.2\mathrm{E}{-62}$ & $1.1\mathrm{E}{-29}$ \\
      & \#iterations & 5 & 6 & 7 & 16 \\
      & seconds & 222 & 807 & 1,952 & 7,579
\end{tabular}
\caption{Newton's method for the power series expansion of a generic point
on a surface of cyclic 128-roots, for truncation degrees 8, 16, 24, and~32,
for quad, penta, octo, and deca double precision.
The seconds record the wall clock time with 88 threads.  }
\label{tabcyc128}
\end{center}
\end{table}

For degree~8, the computations with penta doubles finish in 10~seconds
sooner than the computations with quad doubles, because 5 iterations
suffice.  For degree~16, the results in deca double precision are
much more accurate than in octo double precision, with the same number
of iterations.  Adding up all seconds in Table~\ref{tabcyc128} gives
18,717 seconds, or 5 hours, 11 minutes, and 57~seconds.
Without parallel software, this experiment would have taken 
more than 100 hours, more than 4 days.
The multiplication factor of~20 is derived from the efficiency study
in the next section.
Obviously, parallel software saves time when running numerical experiments.

\subsection*{7. Computational Results}

The runs are done on a CentOS Linux workstation with 256 GB RAM
and two 22-core 2.2 GHz Intel Xeon E5-2699 processors.
If one is mainly interested in the fastest throughput,
then with hyperthreading, runs could be done with 88 threads.
However, the effect of hyperthreading is not equivalent to
doubling the number of cores.  
In the practical evaluation of the parallel implementation,
the runs therefore stop at 40 worker threads.

Random polynomial systems are generated,
64 polynomials with 64 monomials per polynomial.
Power series are truncated to degrees 8, 16, and 32.
Efficiencies are reported for 2, 4, 8, 16, 32, and 40 worker threads.
Efficiency is speedup divided by the number of worker threads.

\begin{figure}[h!]
\begin{center}
{\includegraphics[scale=0.71]{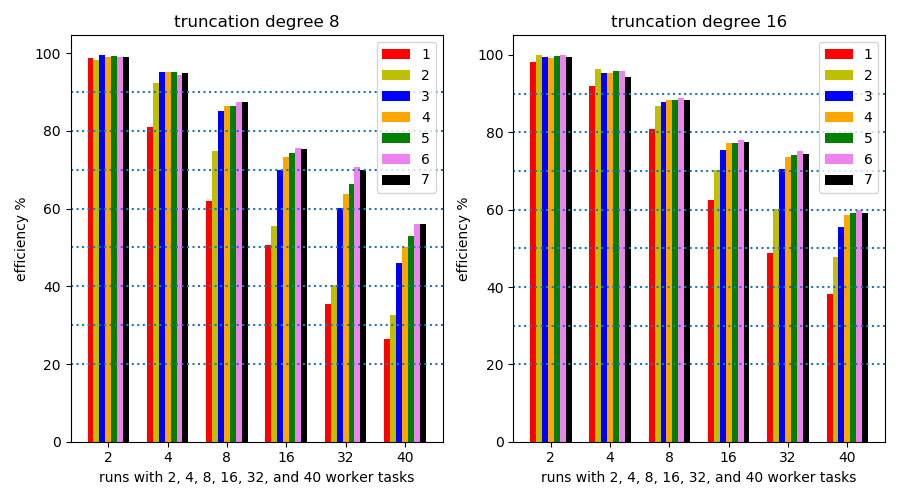}}
\caption{Efficiency plots for power series truncated to degrees 8 and 16,
for 2, 4, 8, 16, 32, and 40 worker tasks, for seven precisions:
double (d), 2-d, 3-d, 4-d, 5-d, 8-d, and 10-d,
in the plots labeled respectively by bars 1, 2, 3, 4, 5, 6, and~7.}
\label{figeffplots1}
\end{center}
\end{figure}

The plots in Figure~\ref{figeffplots1}
show efficiencies for degrees 8 and 16
of the truncated power series.
The efficiencies decrease from close to 100\% 
(a near perfect speedup for 2 threads) to below 60\%
when 40 worker threads are used.
As efficiency equals speedup divided by the number of worker threads,
the speedup corresponding to 60\% efficiency for 40 worker threads 
equals $0.6 \times 40 = 24$.

\begin{figure}[t!]
\begin{center}
{\includegraphics[scale=0.71]{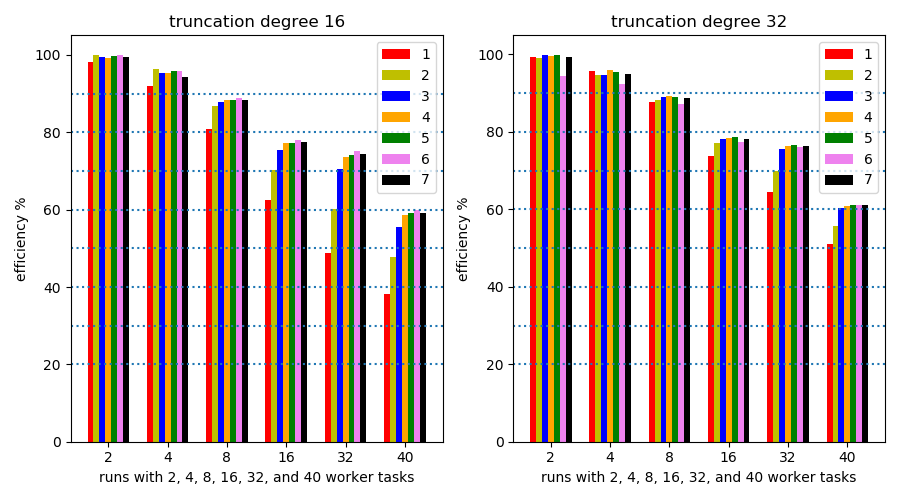}}
\caption{Efficiency plots for power series truncated to degrees 16 and 32,
for 2, 4, 8, 16, 32, and 40 worker tasks, for seven precisions:
double (d), 2-d, 3-d, 4-d, 5-d, 8-d, and 10-d,
in the plots labeled respectively by bars 1, 2, 3, 4, 5, 6, and~7.}
\label{figeffplots2}
\end{center}
\end{figure}

The plots in Figure~\ref{figeffplots2}
compare the efficiencies for degrees 16 and 32.
For truncation degree~32, we observe that 60\% efficiency is
reached already at triple double precision.
More extensive numerical experiments would increase the number
of polynomials and the number of monomials per polynomial to
investigate the notion of isoefficiency.  In particular,
by how much should the size of the problem increase to obtain
the same efficiency as the number of threads increases?

The computational results of this section
(the 60\% efficiency or the 24 speedup) 
justify the multiplication factor of 20 used 
in the last paragraph of section~6.

\newpage

\subsection*{8. Conclusions}

This paper presents a use case of multiple double precision
in the application of Newton's method to develop power series expansions
for solution curves of polynomial systems.
The experiments described in this paper are performed by recent additions
to the free and open source software PHCpack, available via github.

PHCpack contains an Ada version of the code in QDlib~\cite{HLB01},
for double double and quad double precision, and of code generated
by the software CAMPARY~\cite{JMPT16},
for triple, penta, octo, and deca double precision.
The cost overhead factors of multiple double precision are multiplied
with the cost overhead factors of truncated power series arithmetic.
This cost overhead justifies the application of multitasking 
to write parallel software.
Using all kernel threads on a 44-core computer,
numerical experiments that took about 5 hours are estimated to take
more than four days without multitasking.

The efficiency of the current implementation is limited by
the medium grained parallelism and may not scale well
on shared memory computers with over one hundred cores.
In refining the granularity of the current implementation, 
the Ada 202X parallel features look promising.

\noindent {\bf Acknowledgments.}
The author thanks Clyde Roby, Tucker Taft, and Richard Wai,
the organization committee of the HILT 2020 Workshop on Safe Languages
and Technologies for Structured and Efficient Parallel 
and Distributed/Cloud Computing.  
Earlier versions of the software were presented in the Ada devroom 
at FOSDEM 2020.  The author is grateful to Dirk Craeynest and 
Jean-Pierre Rosen for the organization of the FOSDEM 2020 Ada devroom.

\newpage

\bibliographystyle{plain}

\begin{thebibliography}{15}

\bibitem{BG96}
G.~A. Baker and P.~Graves-Morris.
\newblock {\em {Pad{\'{e}} Approximants}}, volume~59 of {\em Encyclopedia of
          Mathematics and its Applications}.
\newblock Second edition, Cambridge University Press, 1996.

\bibitem{BW07}
A. Burns and A. Wellings.
\newblock {\em Concurrent and Real-Time Programming in Ada.}
\newblock Cambridge University Press, 2007.

\bibitem{BV18}
N. Bliss and J. Verschelde.
\newblock The method of Gauss-Newton to compute power series solutions
          of polynomial homotopies.
\newblock {\em Linear Algebra and Its Applications} 542:569--588, 2018. 

\bibitem{GV83}
G.~H. Golub and C.~F. Van Loan.
\newblock {\em Matrix Computations.}
\newblock The Johns Hopkins University Press, 1983.

\bibitem{GW08}
A.~Griewank and A.~Walther.
\newblock {\em Evaluating Derivatives: Principles and Techniques
          of Algorithmic Differentiation}.
\newblock SIAM, 2008.

\bibitem{HLB01}
Y. Hida, X.~S. Li, and D.~H. Bailey.
\newblock Algorithms for quad-double precision floating point arithmetic.
\newblock In the {\em Proceedings  of the 15th IEEE Symposium on Computer
          Arithmetic (Arith-15 2001)}, pages 155--162.
\newblock IEEE Computer Society, 2001.

\bibitem{JMPT16}
M. Joldes, J.-M. Muller, V. Popescu, W. Tucker.
\newblock CAMPARY: Cuda Multiple Precision Arithmetic Library and
          Applications.
\newblock In {\em Mathematical Software -- ICMS 2016, the 5th
          International Conference on Mathematical Software}, pages 232--240,
\newblock Springer-Verlag, 2016.

\bibitem{MSH11}
J.~W. McCormick, F. Singhoff, and J. Hugues.
\newblock {\em Building Parallel, Embedded, and Real-Time Applications 
          with Ada.} 
\newblock Cambridge University Press, 2011.

\bibitem{Mor87}
A.~Morgan.
\newblock {\em Solving Polynomial Systems using Continuation for Engineering
  and Scientific Problems}, volume~57 of {\em Classics in Applied Mathematics}.
\newblock SIAM, 2009.

\bibitem{MBDJJLMRT18}
J.-M. Muller, N. Brunie, F. de Dinechin, C.-P. Jeannerod, M. Joldes,
V. Lefevre, G. Melquiond, N. Revol, S. Torres.
\newblock {\em Handbook of Floating-Point Arithmetic.}
\newblock Second Edition, Springer-Verlag, 2018.

\bibitem{Rum10}
S.~M. Rump.
\newblock Verification methods: Rigorous results using floating-point
          arithmetic.
\newblock {\em Acta Numerica} 19:287--449, 2010.

\bibitem{TVV20a}
S. Telen, M. Van Barel, and J. Verschelde.
\newblock A robust numerical path tracking algorithm 
          for polynomial homotopy continuation.
\newblock {\em SIAM Journal on Scientific Computing}
          42(6):A3610--A3637, 2020.

\bibitem{TVV20b}
S. Telen, M. Van Barel, and J. Verschelde.
\newblock Robust numerical tracking of one path of a polynomial homotopy
          on parallel shared memory computers.
\newblock In the {\em Proceedings of the 22nd International Workshop on 
          Computer Algebra in Scientific Computing (CASC 2020)},
          volume 12291 of {\em Lecture Notes in Computer Science},
          pages 563--582.
\newblock Springer-Verlag, 2020. 

\bibitem{Ver99}
J.~Verschelde.
\newblock Algorithm 795: {PHCpack}: A general-purpose solver for polynomial
          systems by homotopy continuation.
\newblock {\em ACM Transactions on Mathematical Software}
          25(2):251--276, 1999.  Runs online at {\tt www.phcpack.org}.

\bibitem{Wal50}
R.~J. Walker.
\newblock {\em Algebraic Curves.}
\newblock Princeton University Press, 1950.

\end{thebibliography}

\end{document}